\documentstyle[preprint,revtex]{aps}
\draft
\begin{document}
\begin{title}
Critical Exponents for Three-Dimensional \\
Superfluid$-$Bose-Glass Phase Transition
\end{title}
\author{Lizeng Zhang}
\begin{instit}
Department of Physics and
Astronomy, \\
The University of Tennessee, Knoxville,
Tennessee 37996-1200 \\
and Solid State Division, Oak Ridge National Laboratory\\
Oak Ridge, Tennessee 37831
\end{instit}
\author{ Xiao-Qian Wang}
\begin{instit}
Center for Theoretical Studies of Physical Systems, \\
Clark Atlanta University,
Atlanta, Georgia 30314
\end{instit}
\begin{abstract}
The critical phenomenon of the zero temperature
superfluid$-$Bose-glass phase
transition for hard-core bosons
on a three-dimensional disordered lattice
is studied using a quantum real-space renormalization-group
method. The correlation-length exponent $\nu$ and
the dynamic exponent $z$ are computed. The critical exponent
$z$ is found to be 2.5 for compressible states and 1.3 for
incompressible states. The exponent $\nu$
is shown to be insensitive to $z$ as that
in the two-dimensional case, and has value roughly equal to 1.
\end{abstract}

\narrowtext

Disordered boson system provides a prototypical example of
zero temperature quantum critical phenomena \cite{MHL,FWGF}.
Experimentally,
such a disordered boson system can be realized in liquid
He$^{4}$ in random media \cite{EXPT}, or in disordered
superconductors \cite{LIU} where cooper pairs can be
modeled as composite bosons.
As the amount of disorder is varied, these systems exhibit a
continuous
phase transition from the superfluid (SF) phase to a disordered
(Bose-glass (BG)) phase. Understanding of this quantum critical
phenomenon has been a subject of a considerable
amount of recent experimental and theoretical studies.

Besides a diverging length scale (the correlation length) $\xi $,
the SF-BG transition is also characterized by a diverging time scale
$\tau $. Denoting $\delta$ as the distance to the criticality,
$\xi$ and $\tau$ can be described by the critical exponents $\nu$
and $z$, defined by $\xi \propto \delta^{-\nu}$ and $\tau \propto
\xi^{z} \propto \delta^{-\nu z}$. From a general scaling argument,
Fisher {\it et al.}
\cite{FWGF} concluded that for compressible states
$z$ is equal to $d$, the dimensionality
of the system, while $z=1$ for systems with long-range Coulumb
interactions in any dimension \cite{FG}. The critical
exponent $\nu$ cannot be deducted directly from the scaling theory,
but a rigorous lower bound has been established,
i.e., $\nu  \geq 2/d$ \cite{CCFS}.
Unfortunately, standard field-theoretical
renormalization-group (RG) method, which is
proven to be a powerful approach in the study
of critical phenomena, has
been eluded so far from being applied to this SF-BG phase
transition. A major difficulty associated with this approach
is due to the lack of a proper zero loop (mean field) theory
describing the SF-BG transition at finite dimension
\cite{FWGF}, upon which
perturbation series, such as $\epsilon$-expansions, can be developed.
For this reason, a different RG formulation, namely
the real-space RG (RSRG) approach, becomes a
useful alternative for investigating
the critical phenomena in disordered boson systems.

RSRG has been applied previously to one-dimensional (1D)\cite{ZM2,SR},
and two-dimensional (2D) systems \cite{ZM2}. In this paper, we apply
the RSRG method developed previously by Zhang and Ma \cite{ZM2}
to investigate the SF-BG transition in three dimension.
Such a study is important not only because of its intrinsic
theoretical interest, but also due to its relevance to
experiments. Three dimensional (3D) disordered boson systems
are directly realized in liquid He$^{4}$ in Vycor,
aerogels, xerogels, and other random media \cite{EXPT}.
While present experiments are mainly focused on the
effect of disorder on the finite temperature
superfluid phase transition, refined
experiments are hopefully to be capable of
extracting information about the
zero temperature critical point in the near future.
It is clearly desirable to perform calculations
for critical exponents in 3D which characterize the
physical properties of the system in the vicinity of
the transition.

In the following we shall briefly outline the RSRG scheme which has
been presented in detail in Ref. \cite{ZM2}.
The RSRG scheme yields, when applied to the 1D
systems, no (non-trivial) fixed point, indicating the
instability of
the superfluid phase against any amount of disorder for hard-core
boson
systems, in agreement with other 1D RG calculations \cite{GS} and
exact results \cite{ZM1}.
For 2D and 3D systems, it gives a non-trivial fixed point separating
the superfluid phase and the disordered phase.
In 2D, the critical exponent $z$ was found to be
about 1.7 for compressible states and
about 0.9 for incompressible states. Since $z$
is close to 2 and 1 in these two
cases, and since it increases slowly with increasing
block size \cite{ZM2}, this calculation can be viewed as a
supporting evidence of the scaling
prediction by Fisher {\it et al.} \cite{FWGF} in 2D.
The critical exponent $\nu$ is found to be quite insensitive to
the type of states, and is roughly equal to 1.4 \cite{ZM2}.
This value of $\nu$ satisfies
the rigorous lower bound of Chayes {\it et al.} \cite{CCFS},
and is also consistent with other
numerical estimates \cite{RUNGE}.
The product $z\nu$ for incompressible states may be directly
compared with data obtained from experiments on
superconductor-insulator phase transitions in homogeneous
amorphous films \cite{LIU,ZM2}.
For 3D systems, on the other hand, no systematic computation
has yet been performed to
our knowledge. Other numerical methods used for 1D and 2D
systems, such
as exact diagonalization \cite{RUNGE} or quantum Monte Carlo
simulations \cite{MC},
seem quite formidable for 3D disordered systems
within the present computational capacity.
Thus, the RSRG method provides a unique way to probe the
physical properties of the 3D SF-BG transition.

The system under consideration is a
lattice model of hard-core bosons with random
potential \cite{MHL},
\begin{equation}\label{hcb}
{\cal H} = -t\sum_{\langle i,j
\rangle} (b^{\dagger}_{i}b_{j} + H.c.) + \sum_{j}(W_{j}-\mu)
b^{\dagger}_{j}b_{j} \;\;,
\end{equation}
where $b^{\dagger}_{j}$ and $b_{j}$
denote boson creation and annihilation
operators at lattice site $j$. $\langle i,j \rangle$
indicates nearest-neighbor summation
and $W_{j}$ is the random on-site potential with
(independent) Gaussian distribution. The hard-core constraint
is enforced by the requirement that at each site
the occupation number $b^{\dagger}_{j}b_{j}$
equals to either 0 or 1. While it is
clearly a simplification to the
realistic systems, this model is believed to have
captured the essential physics of the zero-temperature
SF-BG phase transition \cite{MHL,FWGF}.

This hard-core boson model is equivalent to a quantum
spin-$\frac{1}{2}$ $XY$-model with
transverse random fields \cite{MHL},
\begin{equation}
{\cal H} = -J\sum_{<i,j>}(S_{i}^{x}S_{j}^{x} +
S_{i}^{y}S_{j}^{y}) - \sum_{j} h_{j} S_{j}^{z} \;\; ,
\end{equation}
via the mapping $ S^{\dagger}_{j} \leftrightarrow b^{\dagger}_{j}$,
$ S^{-}_{j} \leftrightarrow b_{j}$,
$ J \leftrightarrow 2t $, and
$ h_{j} \leftrightarrow \mu - W_{j} $.
These two equivalent representations of
the problem provide a convenient
way to study the underlying physics of the system
\cite{ZM2,ZM3}, and we will use them alternatively throughout
our discussions.

Our real-space RG method consists of the following steps:

$i$) break the lattice into blocks of size $n_{s}$;

$ii$) compute the block spin which is given by two low energy
eigenstates of the block Hamiltonian;
Since $S^{z}\equiv\sum_{j} S_{j}^{z}$
(corresponding to the
particle number $N_{p} \equiv \sum_{j} b_{j}^{\dagger}b_{j}$ in
the boson language) is a good quantum number,
eigenstates of ${\cal H}$ are also
simultaneously eigenstates of $S^{z}$ ($N_{p}$). As described in
Ref. \cite{ZM2}, this can be accomplished
either by ($a$) selecting the
lowest states of two chosen subspaces of particle number $q$ and
$q+1$,
or by ($b$)
selecting the two lowest states among the ground states of the block
Hamiltonian for each subspace of definite number of particles.
The two states chosen in this way are found to have adjacent
particle number $q'$ and $q'+1$. The field acting on the block spin
is given by the energy difference between the two states. In general,
the block fields will follow a different distribution than that of
site
fields. We chose to keep track only the mean $h=\overline{h_{i}}$
and the variance
$\tilde{h}=(\overline{h_{i}^{2}}-\overline{h_{i}}^{2})^{1/2}$
of the renormalized field, and thus map it onto a Gaussian.

$iii$) calculate the effective couplings between the block spins,
which
is given by the nearest-neighbor couplings of the site spins between
two adjacent blocks;
Due to the presence of disorder, couplings between block spins
are also randomized in the RG iterations. However, these block
couplings are found to be always positive, and the
renormalized system remains unfrustrated. This allows one to
approximate
the block couplings by its mean, and thus confine the RG iterations
within the original parameter space.

$iv$) repeat the RG iteration defined above to find fixed point(s)
and compute critical exponents.

Here we compute the exponents $z$ and $\nu$ for 3D systems. For
computational details, see Ref. \cite{ZM2}. In the
procedure ($a$), fluctuations in particle number at any given region
is one. Since the block size is fixed and it allows only a discrete
set number of particles occupying a block, states
described by such a procedure are incompressible.
The procedure ($b$), which is valid only at the `particle-hole
symmetrical' filling (half boson per site in our case) where
density is conserved through statistical fluctuations in the random
fields \cite{ZM2}, yields a compressible ground state.

In the present work we used a simple-cubic lattice with
block size $n_{s} = 2 \times 2 \times 2$.
Gaussian random fields were generated
numerically and typically an average of
$3 \times 10^{3}$ random configurations
was performed. In the previous work \cite{ZM2}, calculations
for 3D systems were carried out only for the incompressible
state with $q=6$.
Here we complete the study for the
incompressible states with other $q$ values, and
investigate the compressible states as well.
Results for fixed points and critical exponents
are summarized in the table. For incompressible states
with other $q$ values, they can be obtained through
the `particle-hole symmetry'\cite{ZM2}.

As anticipated, critical exponents obtained for incompressible
states at different density have roughly the same value,
indicating that the same universality class is probed.
As that in the 2D case, the critical exponent $\nu$ in 3D is
shown to be rather
insensitive to the procedures adopted, roughly equal to 1.0.
The critical exponent $z$, on the other hand, is quite different for
the compressible and incompressible states.
The present RSRG calculation yields
$z=2.5$ and $1.3$ for the
compressible and incompressible states, respectively.

Our results for $\nu $ in both
compressible and incompressible states satisfy the lower bound of
Chayes {\it et al.} \cite{CCFS}, and provide the first systematic
calculation on critical exponents for 3D systems.
General scaling argument suggests that $z=d$ if the system is
compressible \cite{FWGF}. For systems with long range (Coulomb)
force, dimensional analysis indicates that $z=1$ in any spatial
dimension \cite{FG}. If one assumes that
they belong to the same universality class
as those incompressible states considered here, this value should be
compared with our results obtained through the fixed $q$ procedure.
Our RSRG results for $z$ in 3D deviate more from these scaling
predictions than those obtained for the 2D case \cite{ZM2}.
The value of $z$ deviates from the
scaling result by $17\%$ for the compressible states, and by $30\%$
for the incompressible cases.
The corresponding deviation for 2D systems
is $15$ and $10\%$, respectively.
It is worth to remark that within our RG procedure, the
value of $z$ for compressible states differs from that
for incompressible states roughly by a factor of two in both 2D and
3D cases. While in 2D this is exactly what the scaling argument
suggested, it is clearly not the case in 3D. Our
computation is limited to the cubic block of size $n_{s}=8$,
thus we cannot address the question of size effects. The
next isotropic block has the size $n_{s}=3^{3}=27$ for which
the calculation is computationally not feasible for the time being.
Thus we are not able to test the
direction in which the result  would converge with increasing block
size. Since the RSRG calculation is expected to
approach the exact result as the block size increases \cite{ZM2}, such
a finite-size study is highly desirable.
On the other hand, from the study of 2D systems \cite{ZM2}
one expects that the RSRG estimate for critical exponents
itself is rather insensitive to small changes of the block size.

Although the RG procedure (b) for the compressible states
is valid only at the `particle-hole symmetric' point where
particle number is conserved through the statistical fluctuation
in random fields, it nevertheless allows one to probe
the parameter space away from this special point by
studying the flow of the RG iterations.
The RG flow diagram is similar to what one
gets in 2D with even block size $n_{s}$\cite{ZM2}.
Again, the (unstable) fixed
point at $h=0$ axis controls the critical phenomena of the
SF-BG phase transition.

Before closing, we briefly comment about the RSRG approach to
disordered quantum systems. Since the block states are
computed through an {\it isolated} block Hamiltonian, one
would worry about the effect of the (long-range) coherent quantum
fluctuations which may not be properly taken into account.
A procedure to overcome this difficulty
by incorporating different boundary
conditions for the block states has been recently
suggested, and has been applied to
noninteracting fermion systems \cite{WHITE}. While a direct
application
of such a RG method to disordered system has not been possible
within present computational facilities, we would like to
remark that by averaging over random configurations one can
partly achieve the goal that one wishes to accomplish by
averaging over the boundary conditions. Indeed, it seems
quite ironic that our RSRG
procedure works precisely in the `strong' disordered regime
where the effect of fluctuations can be incorporated through
averaging over the randomness, and losses its validity as the
pure limit is approached \cite{ZM2}.

In summary, we have studied a hard-core disordered boson
system in 3D cubic lattice using a quantum RSRG method.
We have found that there exists a non-trivial fixed point describing
the zero temperature SF-BG phase transition. The critical
exponent $z$ for incompressible states is about 1.3, and
about 2.5 for compressible states. The exponent $\nu$
is  insensitive to $z$, and is roughly equal to 1.0.
As in the 2D case, only one universality class is found, and
the critical behavior of the zero temperature SF-BG
phase transition is controlled
by the `particle-hole symmetric' fixed point.

\vspace{2cm}

We thank M. Ma for discussions.
LZ acknowledges support by the National Science Foundation
under Grant No. DMR-9101542 and by the U.S.
Department of Energy through Contract No. DE-AC05-84OR21400
administered by Martin Marietta Energy Systems Inc.;
XQW acknowledges support by the National Science Foundation
under Grant No. HRD-9154077 and by the
Minnesota Supercomputer Institute.

\begin{table}
\begin{tabular}{cccc}
state & $(h/J)^{*}$ & $\nu $ & $z$ \\ \tableline
incompressible ($q=6$)$\dagger$  & 1.6 & 1.0 & 1.2 \\
incompressible ($q=5$)      & 2.4 & 1.0 & 1.3 \\
incompressible ($q=4$)      & 2.8 & 1.0 & 1.3 \\ \tableline
compressible & 6.0 & 1.1 & 2.5 \\
\end{tabular}
\end{table}
$\dagger$ calculated in ref.\cite{ZM2}.

\end{document}